\documentclass[a4paper,fleqn]{article}
\usepackage{modsim}
\usepackage{times}
\usepackage{natbib}
\usepackage{amsmath, amssymb, amsthm}

\MODSIMhead{Adams {\it et al.}, Efficient sampling from a multivariate normal distribution subject to linear constraints}

\usepackage{rotating}
\usepackage{amsbsy,enumerate,bm}
\usepackage{graphicx}
\usepackage{ccaption}
\usepackage{xcolor}
\usepackage[allbordercolors=white]{hyperref}
\newcommand{\orcid}[1] {\hspace*{-1.5mm} \href{https://orcid.org/#1}{\includegraphics[scale=1.0]{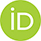}}}

\makeatletter
\renewcommand{\fnum@figure}[1]{\textbf{\figurename~\thefigure}. }
\renewcommand{\fnum@table}[1]{\textbf{\tablename~\thetable}. }
\makeatother

\newcommand{\aand}{\textsuperscript{,} }

\begin{document}

\title{Efficient sampling from a multivariate normal distribution subject to linear equality and inequality constraints}

\author{
\underline{M. P. Adams}
	\address[QUT_SMS]{\it{School of Mathematical Sciences and Centre for Data Science, Queensland University of Technology, Brisbane, Queensland, 4001}}\aand
	\address[QUT_SAEF]{\it{Securing Antarctica's Environmental Future, Queensland University of Technology, Brisbane, \mbox{Queensland, 4001}}}
	 \orcid{0000-0003-4875-0225},
G. M. Monsalve-Bravo
	\address[UQ_ChemEng]{\it{School of Chemical Engineering, The University of Queensland, St Lucia, Queensland, 4072}}	
	 \orcid{0000-0002-6693-4899},
L. G. Dowdell
	\addressmark[QUT_SMS]\aand\addressmark[QUT_SAEF]
	\orcid{0009-0006-9946-9707}, \\
S. A. Sisson
	\address[UNSW_Maths]{\it{School of Mathematics and Statistics and UNSW Data Science Hub, University of New South Wales, Sydney, New South Wales, 2052}}
	\orcid{0000-0001-8943-067X},
and C. Drovandi
 	\addressmark[QUT_SMS]\aand
 	\address[QUT_MACSYS]{\it{Centre of Excellence for the Mathematical Analysis of Cellular Systems, Queensland University of Technology, Brisbane, Queensland, 4001}}
 	\orcid{0000-0001-9222-8763}
}

\email{mp.adams@qut.edu.au}

\date{August 2025}


\begin{abstract}

Sampling from multivariate normal distributions, subjected to a variety of restrictions, is a problem that is recurrent in statistics and computing. In the present work, we demonstrate a general framework to efficiently sample a multivariate normal distribution subject to any set of linear inequality constraints and/or linear equality constraints simultaneously. In the approach we detail, sampling a multivariate random variable from the domain formed by the intersection of linear constraints proceeds via a combination of elliptical slice sampling to address the inequality constraints, and linear mapping to address the equality constraints. We also detail a linear programming method for finding an initial sample on the linearly constrained domain; such a method is critical for sampling problems where the domain has small probability.

We demonstrate the validity of our methods on an arbitrarily chosen four-dimensional multivariate normal distribution subject to five inequality constraints and/or two equality constraints. Our approach compares favourably to direct sampling and/or accept-reject sampling methods; the latter methods vary widely in their efficiency, whereas the methods in the present work are rejection-free. Where practical we compare predictions of probability density functions between our sampling methods and analytical computation. For all simulations we demonstrate that our methods yield accurate computation of the mean and covariance of the multivariate normal distributions restricted by the imposed linear constraints. MATLAB codes to implement our methods are readily available at \href{https://dx.doi.org/10.6084/m9.figshare.29956304}{\color{blue} \underline{https://dx.doi.org/10.6084/m9.figshare.29956304}}.
\end{abstract}

\maketitle

\section{INTRODUCTION}

Sampling from multivariate normal distributions (MNDs), subjected to a variety of restrictions, is a problem that is recurrent in statistics and computing \citep{Robert1995,Lan2023}. After samples are obtained, expectations such as mean vectors and covariance matrices of these distributions can be immediately estimated (e.g.\ \citealt{Adams2022}). Although a wide range of restrictions have been considered in the literature (e.g.\ \citealt{Damien2001,Pakman2014,Maatouk2025}), linear constraints are commonly of interest. Thus, efficient algorithms already exist for sampling MNDs subject to linear equality constraints \citep{Cong2017,Maatouk2022} and linear inequality constraints \citep{Gessner2020,Wu2024,Amrouche2025}. However, we are not aware of any work that demonstrates how to efficiently sample MNDs subject to both linear equality constraints and linear inequality constraints simultaneously.

The domain formed by the intersection of linear inequality constraints can have small probability (so-called ``rare event simulation'', e.g.\ \citealt{Ahn2025}), so a related issue for methods that sample on such a domain is to find one initial sample to begin with. Sampling from MNDs subject to inequality constraints typically rely upon Markov chain Monte Carlo (MCMC) methods such as Gibbs sampling \citep{Chopin2011,Li2015}, elliptical slice sampling \citep{Murray2010,Hasenpflug2025}, or Hamilton Monte Carlo \citep{Pakman2014,Lan2023} -- all of these MCMC sampling methods require an initial sample that is already on the domain of interest. We point out here that for linear constraints, regardless of whether they are inequalities or equalities, an initial sample on a domain satisfying all such constraints can be readily obtained by treating this initial sampling problem as a linear programming model whose optimal solution(s) are \textit{any} samples on the domain. Solution methods for linear programming models are readily available and very fast (e.g.\ \citealt{Hillier2014}).

The present work demonstrates a general framework for efficient sampling of MNDs subject to any set of linear inequality constraints and/or linear equality constraints; the precise problem being solved by this paper is mathematically defined in Section \ref{sec:problem_statement}. Our approach (fully explained in Section \ref{sec:methods}) unifies pre-existing work on sampling MNDs truncated on hyperplanes \citep{Cong2017} and under linear inequality constraints \citep{Gessner2020}. Our approach also ensures that an initial sample is rapidly obtained on the domain of interest using linear programming \citep{Hillier2014}; this is a necessity especially for domains with small probability, but appears to have been overlooked by previous methods. We demonstrate our methods on an arbitrarily chosen four-dimensional MND subject to five inequality constraints and/or two equality constraints (Section \ref{sec:results}). MATLAB codes to implement our methods are readily available at \href{https://dx.doi.org/10.6084/m9.figshare.29956304}{\color{blue} \underline{https://dx.doi.org/10.6084/m9.figshare.29956304}}.

\section{PROBLEM STATEMENT AND THE TRANSFORMED SAMPLING PROBLEM}
\label{sec:problem_statement}

We consider the problem of sampling a multivariate, $n$-dimensional, normal variable $\bm{x} \in \mathbb{R}^n$ subject to both linear inequality constraints and linear equality constraints,
\begin{equation}
\bm{x} \sim \mathcal{N}(\bm{\mu},\Sigma), \quad \mbox{subject to} \quad A \bm{x} + \bm{b} \geq \bm{0} \quad \mbox{and} \quad C \bm{x} + \bm{d} = \bm{0},
\label{eq:problem_definition}
\end{equation}
where the MND is characterised by a mean vector $\bm{\mu} \in \mathbb{R}^n$ and symmetric and positive semi-definite covariance matrix $\Sigma \in \mathbb{R}^{n \times n}$, there are $m \geq 0$ linear inequality constraints as defined by matrix $A \in \mathbb{R}^{m \times n}$ and vector $\bm{b} \in \mathbb{R}^m$, and there are $p \geq 0$ linear equality constraints defined by matrix $C \in \mathbb{R}^{p \times n}$ and vector $\bm{d} \in \mathbb{R}^p$. 

Using Algorithm 2 of \cite{Cong2017}, together with a coordinate transformation, we can convert this to a problem of sampling a latent variable $\bm{y} \in \mathbb{R}^n$ from a MND with zero mean but the same covariance $\Sigma$, so that $\bm{x}$ can be obtained from $\bm{y}$ via the linear mapping
\begin{equation}
\bm{x} = F \bm{y} + \bm{g},
\label{eq:recast_x}
\end{equation}
where the matrix $F \in \mathbb{R}^{n \times n}$ and the vector $\bm{g} \in \mathbb{R}^n$. Using this linear mapping, the problem can be reduced to sampling $\bm{y}$ from only a set of linear inequality constraints,
\begin{equation}
\bm{y} \sim \mathcal{N}(\bm{0},\Sigma),
\label{eq:recast_y}
\end{equation}
subject to
\begin{equation}
H \bm{y} + \bm{k} \geq \bm{0},
\label{eq:recast_inequality}
\end{equation}
where the matrix $H \in \mathbb{R}^{m \times n}$ and the vector $\bm{k} \in \mathbb{R}^m$.

Transformation of the sampling problem of equation \eqref{eq:problem_definition} to the sampling problem described by equations \eqref{eq:recast_x}-\eqref{eq:recast_inequality} can be achieved by computing $F$, $\bm{g}$, $H$ and $\bm{k}$ (as well as computing an intermediate matrix $E$), from $A$, $\bm{b}$, $C$, $\bm{d}$, $\bm{\mu}$ and $\Sigma$, via
\begin{align}
\label{eq:conversion_E} E &= \Sigma C^\top \left( C \Sigma C^\top \right)^{-1},\\
\label{eq:conversion_F} F &= I - EC, \\
\label{eq:conversion_g} \bm{g} &= F \bm{\mu} - E \bm{d}, \\
\label{eq:conversion_H} H &= AF, \\
\label{eq:conversion_k} \bm{k} &= A \bm{g} + \bm{b},
\end{align}
where $I$ is the identity matrix of size $n \times n$. 

\section{METHODS}
\label{sec:methods}

The methods needed for the sampling problem given by equation \eqref{eq:problem_definition} depend on whether the variable $\bm{x} \sim \mathcal{N}(\bm{\mu},\Sigma)$ is subject to only linear equality constraints, only linear inequality constraints, or both linear equality constraints and linear inequality constraints. In the following subsections, we detail general methods for each of these three different problems.

\subsection{Efficient sampling from a MND subject to linear equality constraints}
\label{sec:algorithm_case_equality_only}

In this case, the sampling problem given by equation \eqref{eq:problem_definition} requires definition of $\bm{\mu}$, $\Sigma$, $C$ and $\bm{d}$, but does not require definition of $A$ or $\bm{b}$ (since there are no linear inequality constraints present).

The sampling can proceed via the following steps:
\begin{enumerate}
\item Establish if the linear system $C \bm{x} + \bm{d} = \bm{0}$ has zero, one or infinite solutions for $\bm{x}$. If the linear system has zero solutions, the sampling problem is impossible, and the algorithm concludes unsuccessfully. If the linear system has one solution $\bm{x}$, all samples of this system will be equal to this one value of $\bm{x}$ (i.e.\ there is a point mass at $\bm{x}$), and the algorithm concludes successfully (but trivially). If the linear system has infinite solutions, proceed to Step 2.
\item Compute $E$, $F$ and $\bm{g}$ according to equations \eqref{eq:conversion_E}-\eqref{eq:conversion_g}.
\item Sample the latent variable $\bm{y}$ according to equation \eqref{eq:recast_y}, and thereafter obtain samples $\bm{x}$ from equation \eqref{eq:recast_x}.
\end{enumerate}
This procedure is very fast, and is mathematically equivalent to Algorithm 2 of \cite{Cong2017}.

\subsection{Efficient sampling from a MND subject to linear inequality constraints}
\label{sec:algorithm_case_inequality_only}

In this case, the sampling problem given by equation \eqref{eq:problem_definition} requires definition of $\bm{\mu}$, $\Sigma$, $A$ and $\bm{b}$, but does not require definition of $C$ or $\bm{d}$ (since there are no linear equality constraints present).

The sampling can proceed via the following steps:
\begin{enumerate}
\item Do not calculate the intermediate matrix $E$. Instead, compute $F$, $\bm{g}$, $H$ and $\bm{k}$ as follows:
\begin{equation}
F = I, \quad \bm{g} = \bm{\mu}, \quad H = A, \quad \bm{k} = A \bm{\mu} + \bm{b}.
\label{eq:simple_FGHK}
\end{equation}
Combining equations \eqref{eq:recast_x} and \eqref{eq:simple_FGHK} implies that any computation of samples of $\bm{x}$ from latent variable samples $\bm{y}$ (if required later in the algorithm) will simply be
\begin{equation}
\bm{x} = \bm{y} + \bm{\mu}.
\label{eq:simple_x}
\end{equation}
\item Establish if there are zero, one or infinite values of $\bm{y}$ that satisfy $H \bm{y} + \bm{k} \geq \bm{0}$. This can be accomplished by solution of the following linear programming problem that aims to minimise a defined objective value $z$:
\begin{equation}
\min_{\bm{y},\bm{a}} (z) = \min_{\bm{y},\bm{a}} \left(\displaystyle\sum_i a_i \right) \quad \mbox{subject to} \quad H \bm{y} + \bm{k} + \bm{a} \geq \bm{0}, \quad \mbox{and} \quad \bm{a} \geq \bm{0}.
\label{eq:LP_for_z0}
\end{equation}
Note here that $\bm{a} \in \mathbb{R}^m$ should be understood as being a column vector of the same length as $\bm{k}$, with elements $a_i$ where $i=1,...,m$, and the elements of the decision variable vector $\bm{y}$ can take any real value.

If the optimal value of the objective function $z$ in equation \eqref{eq:LP_for_z0} is positive, then there are no values of $\bm{y}$ that satisfy $H \bm{y} + \bm{k} \geq \bm{0}$, so the original sampling problem in equation \eqref{eq:problem_definition} is impossible, and the algorithm concludes unsuccessfully. If the optimal value of the objective $z$ in equation \eqref{eq:LP_for_z0} is zero, then there are either one or infinite vectors $\bm{y}$ that satisfy $H \bm{y} + \bm{k} \geq \bm{0}$, so proceed to Step 3.

\item For the linear programming problem defined in Step 2, if there is only one optimal value of the decision variables $\bm{y}$ for which $z=0$, then all latent variable samples of this system will be equal to this $\bm{y}$, and the algorithm concludes successfully (but trivially). (Samples of $\bm{x}$ are subsequently computed from samples of these latent variables $\bm{y}$ according to equation \eqref{eq:simple_x}.) On the other hand, if there are multiple optimal values of the decision variables $\bm{y}$ for which $z=0$, then any one of these $\bm{y}$ is chosen as the first sample of $\bm{y}$, and subsequently proceed to Step 4.

\item Rejection-free samples of the latent variable $\bm{y}$, which is a multivariate normal variable with zero mean (equation \eqref{eq:recast_y}) subject to linear inequality constraints (inequality \eqref{eq:recast_inequality}) can be obtained via the LIN-ESS algorithm, introduced in \cite{Gessner2020} and modified in \cite{Adams2022}. The LIN-ESS algorithm is a specific type of elliptical slice sampling \citep{Murray2010} that is tailored for rejection-free sampling of a multivariate normal variable from a spatial domain truncated within linear inequalities. Once a sufficient number of samples of the latent variable $\bm{y}$ are obtained, samples of the required variable $\bm{x}$ are subsequently computed from equation \eqref{eq:simple_x}.
\end{enumerate}

For the fourth step, note that LIN-ESS is an MCMC method, so there may be consideration given to discard some samples to account for MCMC burn-in and/or autocorrelation of samples, depending on whether a larger number of less independent samples or a smaller number of more independent samples is desired for the application of interest.  In the present work, we will not discard any samples; for the interested reader, discarding samples from LIN-ESS is also explored in \citet{Adams2022}.

\subsection{Efficient sampling from a MND subject to both linear equality and inequality constraints}
\label{sec:algorithm_case_equality_and_inequality}

In this case, the sampling problem given by equation \eqref{eq:problem_definition} requires definition of $\bm{\mu}$, $\Sigma$, $A$, $\bm{b}$, $C$ and $\bm{d}$.
The sampling can proceed as follows:
\begin{enumerate}
\item Compute $E$, $F$, $\bm{g}$, $H$ and $\bm{k}$ via equations \eqref{eq:conversion_E}-\eqref{eq:conversion_k}. This means that any computation of samples of $\bm{x}$ from latent variable samples $\bm{y}$ (if required later in the algorithm) must proceed via equation \eqref{eq:recast_x}.
\item Follow Steps 2-4 of Section \ref{sec:algorithm_case_inequality_only}, noting that samples $\bm{x}$ are obtained from samples of the latent variable $\bm{y}$ via equation \eqref{eq:recast_x} instead of equation \eqref{eq:simple_x}.
\end{enumerate}

\section{RESULTS AND DISCUSSION}
\label{sec:results}

In what follows, we demonstrate our results on an arbitrarily chosen MND of dimension $n=4$. We introduce a set of linear inequality constraints and a set of linear equality constraints for this MND, and examine three sampling problems: (1) sampling the MND subject to only the inequality constraints, (2) sampling the MND subject to only the equality constraints, and (3) sampling the MND subject to both the inequality and equality constraints.

For each problem, we will compare our sampling method to an appropriate direct or accept-reject sampling method (treated as an approximate ``ground truth'' for our sampling methods). We generate $10^6$ samples from each sampling method. For each problem, the elements of mean vectors and covariance matrices of the samples, denoted as $(\mu_{\tau})_i$ and $(\sigma_{\tau})_{ij}$ respectively, are approximated from and compared between each method to test the validity of our methods. Where relevant and convenient, we also visually inspect the samples, and compare other relevant quantities including estimation of probability density functions via kernel density estimation and analytical computation.

\begin{figure}[ht!] \centering
\includegraphics[width=\textwidth]{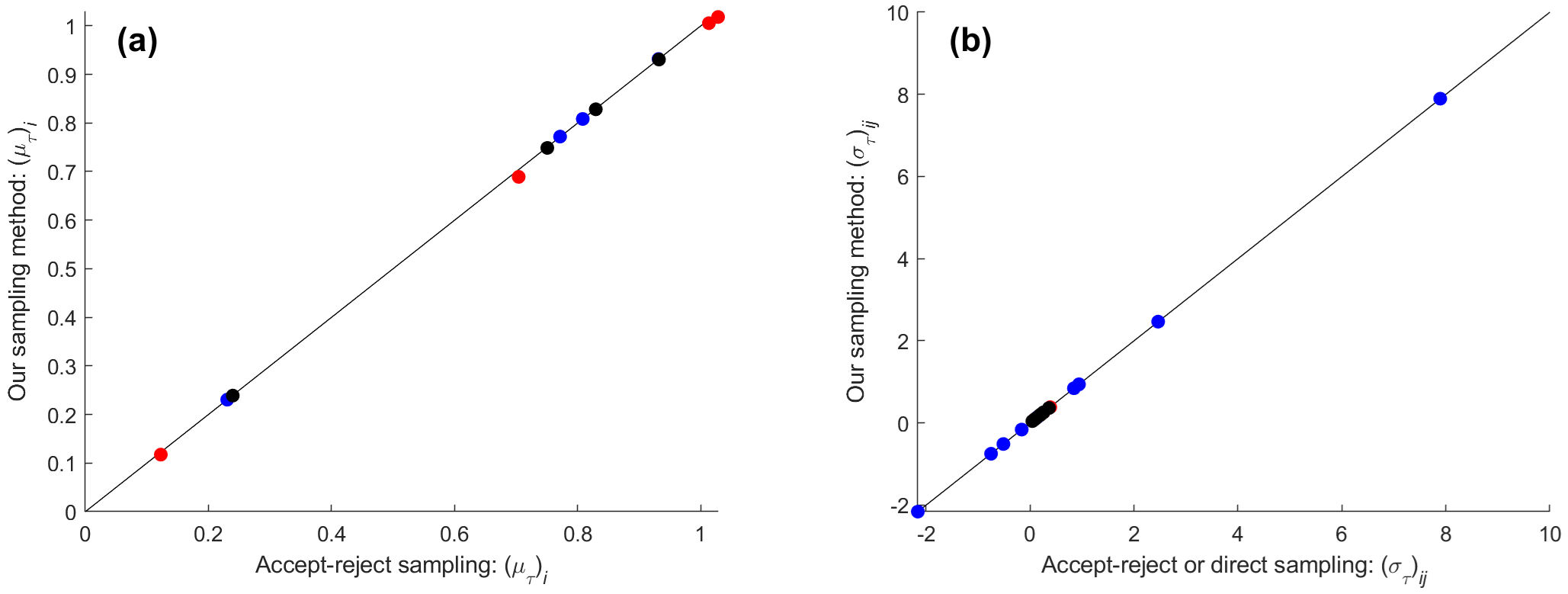}
\caption{Comparison of estimates of mean vector elements $(\mu_\tau)_i$ and covariance matrix elements $(\sigma_\tau)_{ij}$ between our method and either accept-reject sampling or direct sampling (where relevant), for the four-dimensional MND in equation \eqref{eq:pentagon_x} subjected to inequality constraints in \eqref{eq:pentagon_inequal} (red dots), equality constraints in \eqref{eq:pentagon_equal} (blue dots), or both these inequality and equality constraints in (\ref{eq:pentagon_inequal},\ref{eq:pentagon_equal}) simultaneously (black dots). Notice that all points coincide with the 1:1 line (black line), indicating good agreement between sampling method estimates of mean and covariance quantities.}
\label{fig:Pentagon_Mean_Covariance}
\end{figure}

The arbitrarily chosen four-dimensional MND used as an example of our methods in this paper is
\begin{equation}
 \bm{x} \sim \mathcal{N} \left( \begin{bmatrix} 0.284 \\ 0.964 \\ 0.940 \\ 0.664 \end{bmatrix}, \left[\begin{array}{rrrr} 0.960 & 1.407 & 0.754 & -1.360 \\
1.407 &  8.250 & 1.105 & -1.993 \\
0.754 &  1.105 & 14.79 & -7.116 \\
-1.360 & -1.993 & -7.116 & 5.350 \end{array}\right] \right).
\label{eq:pentagon_x}
\end{equation}
We consider this MND to be subject to five arbitrarily chosen inequality constraints,
\begin{equation}
 A \bm{x} + \bm{b} \geq \bm{0}, \quad \mbox{where} \quad
A = \left[\begin{array}{rrrr} 128.61 & 935.51 & -425.89 & -472.28 \\
                    -15.34 & -223.32 & 27.84 & 196.12 \\
                    103.19 & -107.39 & 23.58 & 19.79 \\
                   -923.53 & -5030.49 & 2283.68 & 2670.02 \\
                     83.29 & 466.44 & -204.57 & -254.19 \end{array}\right], \quad
\bm{b} = \left[\begin{array}{rrrr} -183.90 \\ 72.14 \\ 102.45 \\ 1010.98 \\ -83.47 \end{array}\right],
\label{eq:pentagon_inequal}
\end{equation}
and/or two arbitrarily chosen equality constraints (whose intersection yields a two-dimensional plane in four-dimensional space),
\begin{equation}
C \bm{x} + \bm{d} = \bm{0}, \quad \mbox{where} \quad
C = \left[\begin{array}{rrrr} 13.04 & 60.57 & -26.93 & -33.82 \\ 0.36 & -9.15 & 4.00 & 4.05 \end{array}\right], \quad \bm{d} = \left[\begin{array}{rrrr} -11.31 \\ 2.08 \end{array}\right].
\label{eq:pentagon_equal}
\end{equation}
We now explore the validity of our methods for this distribution and group of constraints. First, for sampling the variable $\bm{x}$ in equation \eqref{eq:pentagon_x} subject only to the linear inequality constraints in \eqref{eq:pentagon_inequal}, it is difficult to visualise this problem as it is a four-dimensional space. Instead, we compare our sampling method to accept-reject sampling; for the present problem accept-reject sampling is highly inefficient (acceptance rate of $\approx$ 0.005\%) albeit not completely impractical. If $10^6$ samples are obtained from each sampling method, mean vector elements $(\mu_\tau)_i$ and covariance matrix elements $(\sigma_\tau)_{ij}$ calculated using our samples agree well with accept-reject sampling (red dots in Figure \ref{fig:Pentagon_Mean_Covariance}).

\begin{figure}[ht!] \centering
\includegraphics[width=0.5\textwidth]{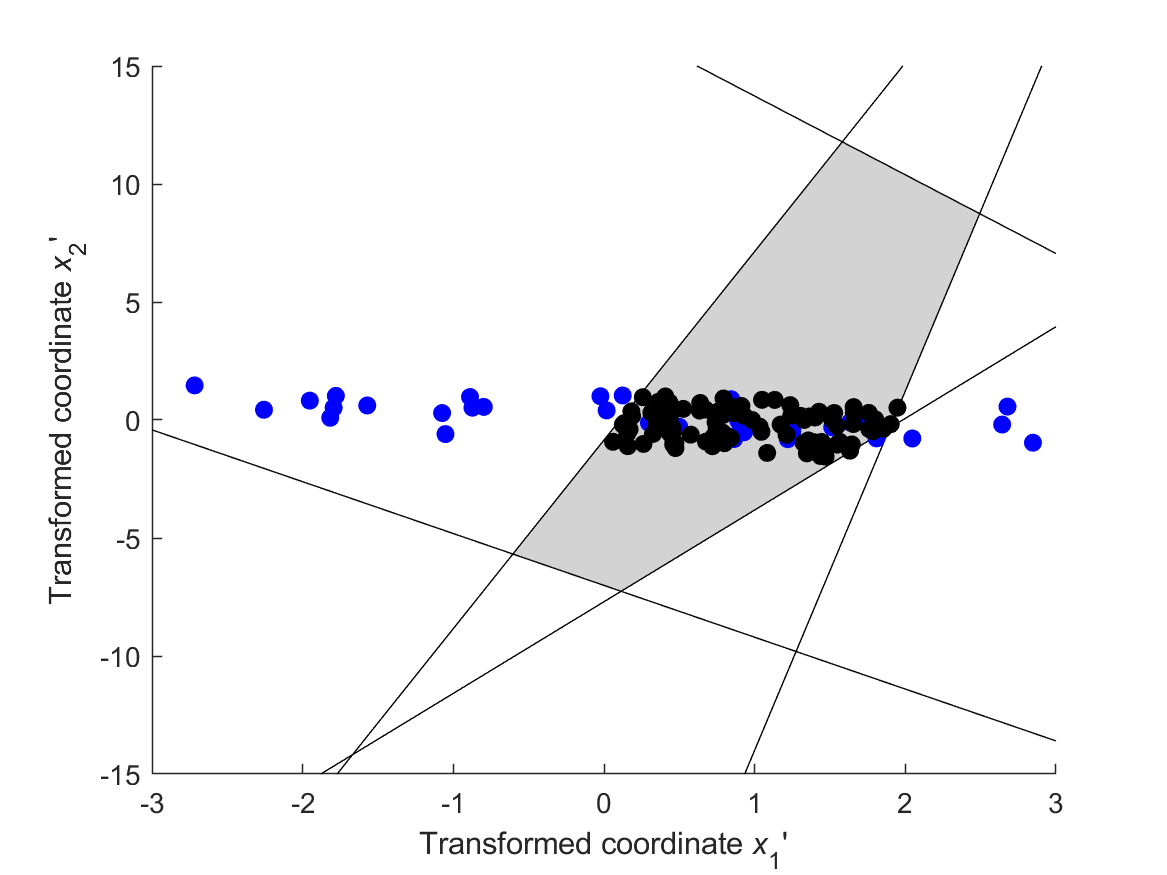}
\caption{Visualisation of the two-dimensional plane, formed by the linear equality constraints in \eqref{eq:pentagon_equal}, on which the four-dimensional MND in equation \eqref{eq:pentagon_x}, is sampled. Lines show the bounds of the linear inequality constraints in \eqref{eq:pentagon_inequal} graphed in this two-dimensional plane ($x_1',x_2'$) whose coordinates are obtained from equation \eqref{eq:pentagon_x_dash}. Blue dots show 100 samples of the MND subject to linear equality constraints in \eqref{eq:pentagon_equal}, obtained from our sampling methods. Black dots show 100 samples of the MND subject to both linear equality constraints in \eqref{eq:pentagon_equal} and linear inequality constraints in \eqref{eq:pentagon_inequal}, obtained from our sampling methods.}
\label{fig:Pentagon_Shape}
\end{figure}

\begin{figure}[ht!] \centering
\includegraphics[width=\textwidth]{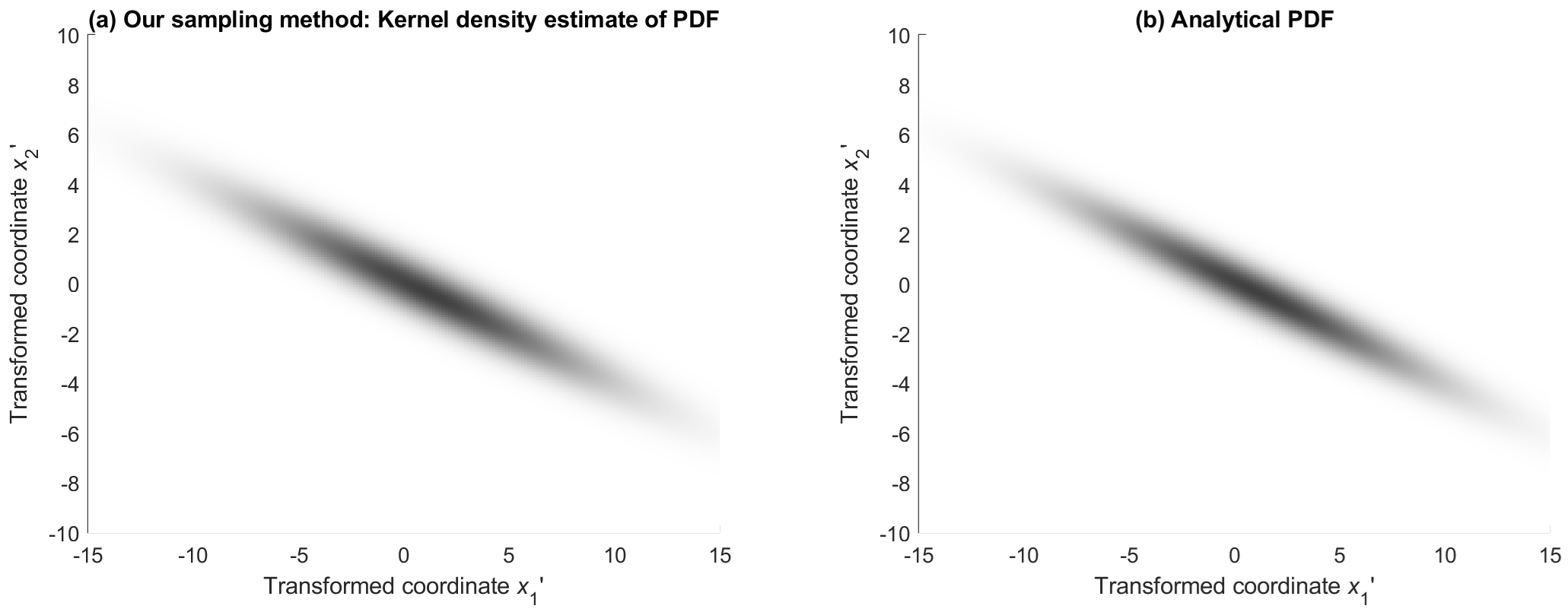}
\caption{Comparison of the probability density function (PDF) predicted by our sampling method (a, left) to an analytical approach (b, right) for the MND in equation \eqref{eq:pentagon_x} subject to the linear inequality constraints in \eqref{eq:pentagon_equal}, when visualised in the $(x_1',x_2')$ plane obtained from the coordinate transformation described in equation \eqref{eq:pentagon_x_dash}. Notice that there is strong agreement between the computed PDFs.}
\label{fig:Pentagon_PDF}
\end{figure}

Second, for sampling the variable $\bm{x}$ in equation \eqref{eq:pentagon_x} subject only to the linear equality constraints in \eqref{eq:pentagon_equal}, it is possible to transform this sampling problem via coordinate transform $\bm{x} \rightarrow \bm{x}'$ of the form
\begin{equation}
\label{eq:pentagon_x_dash}
 \bm{x}' = T \bm{x} + \bm{k}, \quad \mbox{where} \quad T = \left[\begin{array}{rrrr}
-2.25 & -10.68 & 6.24 & 4.21 \\
-8.57 & -33.44 & 14.64 & 21.10 \\
13.04 & 60.57 & -26.93 & -33.82 \\
0.36 & -9.15 & 4.00 & 4.05 \end{array}\right], \quad \bm{k} = \left[\begin{array}{rrrr} 3.28 \\  4.49 \\
 -11.31 \\ 2.08 \end{array}\right],
 \end{equation}
to another four-dimensional MND $\bm{x}' \sim \mathcal{N}(\bm{\mu}',\Sigma')$ for which the inequality constraint $C\bm{x}+\bm{d}=\bm{0}$ transforms conveniently to $x_3' = x_4' = 0$. This means that the equality constraint $C\bm{x}+\bm{d}=\bm{0}$ represents a two-dimensional plane along the $x_1'$ and $x_2'$ axes. Using well-known conditional probability formulae, it is then straightforward to sample $x_1'$ and $x_2'$ from an appropriately derived bivariate MND on this two-dimensional plane and thereafter directly obtain independent samples of $ \bm{x} \sim \mathcal{N}(\bm{\mu},\Sigma)$ subject to $C \bm{x} + \bm{d} = \bm{0}$ without rejection, via $\bm{x} = T^{-1} (\bm{x}' - \bm{k})$. We compare this direct sampling procedure to our sampling method (the latter visualised as blue dots in Figure \ref{fig:Pentagon_Shape}), and we find that the mean vector elements and covariance matrix elements computed from these sampling techniques agree well (blue dots in Figure \ref{fig:Pentagon_Mean_Covariance}). In additional, kernel density estimation of the probability density from our samples (Figure \ref{fig:Pentagon_PDF}a) in the $(x_1',x_2')$ plane agrees visually closely with analytical computation of this probability density from the bivariate MND $\bm{x}' \sim \mathcal{N}(\bm{\mu}',\Sigma')$ conditional on $x_3'=x_4'=0$ (Figure \ref{fig:Pentagon_PDF}b).

Finally, for sampling the variable $\bm{x}$ in equation \eqref{eq:pentagon_x} subject to both the inequality constraints in \eqref{eq:pentagon_inequal} and equality constraints in \eqref{eq:pentagon_equal}, this is equivalent to sampling $\bm{x}$ within an irregular pentagon in the ($x_1',x_2'$)-plane (see grey shaded region in Figure \ref{fig:Pentagon_Shape}). We used accept-reject sampling on the bivariate MND $\bm{x}' \sim \mathcal{N}(\bm{\mu}',\Sigma')$ conditional on $x_3'=x_4'=0$ with acceptance only when the inequality constraints in \eqref{eq:pentagon_inequal} were satisfied, to compare with our sampling method. This accept-reject method was reasonably efficient ($\approx$ 12\% acceptance rate). In addition to visual confirmation that our method is working correctly (black dots in Figure \ref{fig:Pentagon_Shape}), the mean vector elements and covariance matrix elements computed from our sampling method, and the aforementioned accept-reject sampling method, agree well (black dots in Figure \ref{fig:Pentagon_Mean_Covariance}) when $10^6$ samples were obtained from each method.

\section{CONCLUSION}

In the present work, we have introduced methods to efficiently sample MNDs subject to any set of linear inequality constraints and/or linear equality constraints, and we have demonstrated their utility on an arbitrarily chosen four-dimensional MND. Our general framework unifies previous works on this topic \citep{Cong2017,Gessner2020}, and also advantageously ensures the first sample is on the domain of interest, via solution of a suitably chosen linear programming model \citep{Hillier2014}. It is hoped that these methods might find application in a wide variety of problems in statistics and computing \citep{Lan2023}.

\section*{Acknowledgements}

M.\ P.\ Adams and L.\ G.\ Dowdell acknowledge funding from the ARC SRIEAS Grant SR200100005 Securing Antarctica's Environmental Future. C.\ Drovandi acknowledges funding from an ARC Future Fellowship FT210100260. M.\ P.\ Adams thanks B.\ A.\ J.\ Lawson and D.\ J.\ Warne for fruitful discussions during manuscript development. 

\righthyphenmin=100
\bibliography{ms}
\bibliographystyle{agsm}

\end{document}